\documentclass[10pt,letterpaper,superscriptaddress]{revtex4} 
\setcitestyle{super}
\usepackage{amsmath}
\usepackage{graphicx}

\begin{document}


\title{High-Speed Coherent Raman Fingerprint Imaging of Biological Tissues}

\author{Charles H. Camp Jr.}
\author{Young Jong Lee}
\affiliation{Biosystems and Biomaterials Division, National Institute of Standards and Technology, 100 Bureau Dr, Gaithersburg, MD 20899, USA}
\author{John M. Heddleston}
\affiliation{Semiconductor and Dimensional Metrology Division, National Institute of Standards and Technology, 100 Bureau Dr, Gaithersburg, MD 20899, USA}
\author{Christopher M. Hartshorn}
\affiliation{Biosystems and Biomaterials Division, National Institute of Standards and Technology, 100 Bureau Dr, Gaithersburg, MD 20899, USA}
\author{Angela R. Hight Walker}
\affiliation{Semiconductor and Dimensional Metrology Division, National Institute of Standards and Technology, 100 Bureau Dr, Gaithersburg, MD 20899, USA}
\author{Jeremy N. Rich}
\affiliation{Department of Stem Cell Biology and Regenerative Medicine, Cleveland Clinic, 9500 Euclid Ave, Cleveland, OH 44195, USA}
\author{Justin D. Lathia}
\affiliation{Department of Cellular and Molecular Medicine, Cleveland Clinic, 9500 Euclid Ave, Cleveland, OH 44195, USA}
\author{Marcus T. Cicerone}
\email{cicerone@nist.gov}
\affiliation{Biosystems and Biomaterials Division, National Institute of Standards and Technology, 100 Bureau Dr, Gaithersburg, MD 20899, USA}

\begin{abstract}
We have developed a coherent Raman imaging platform using broadband coherent anti-Stokes Raman scattering (BCARS) that provides an unprecedented combination of speed, sensitivity, and spectral breadth. The system utilizes a unique configuration of laser sources that probes the Raman spectrum over 3,000 cm$^{-1}$ and generates an especially strong response in the typically weak Raman ``fingerprint" region through heterodyne amplification of the anti-Stokes photons with a large nonresonant background (NRB) while maintaining high spectral resolution of $<$ 13 cm$^{-1}$. For histology and pathology, this system shows promise in highlighting major tissue components in a non-destructive, label-free manner. We demonstrate high-speed chemical imaging in two- and three-dimensional views of healthy murine liver and pancreas tissues and interfaces between xenograft brain tumors and the surrounding healthy brain matter.
\end{abstract}
\maketitle

Raman spectroscopy is a powerful technique for analyzing the chemical species within biological samples by probing the intrinsic molecular Raman vibrational energy levels. This chemical analysis technique has been extensively applied to a variety of tissue types and pathologies. Among these applications, it has been used to detect neoplasms of varying grade from tissues of the lung \cite{Huang2003}, breast \cite{Haka2005}, and skin \cite{Gniadecka2004} with a high degree sensitivity and specificity. In these studies and others like them \cite{Meyer2011,Kirsch2010,Krafft2006,Koljenovic2002,Nijssen2002}, multiple peaks in the weakly scattering spectral fingerprint region ($<$ 1,800 cm$^{-1}$) are used to discriminate subtly different states of cells and tissues. Historically, the required quality of Raman fingerprint information has been available only with spontaneous Raman scattering, where spectral collection times are typically in the range of 0.2 s to 30 s\cite{Meyer2011,Kirsch2010,Krafft2006,Koljenovic2002,Nijssen2002}, seriously limiting its use in large-area, high-resolution imaging that is critical for widespread adoption in biological research and clinical practice. To bolster the inherently weak Raman scattering process, coherent Raman imaging (CRI) techniques have been developed that coherently populate selected vibrational states of molecules through their nonlinear response to multiple pulsed laser fields. 

CRI techniques were reintroduced in the late 1990's with the development of coherent anti-Stokes Raman scattering (CARS) microscopy using co-linear signal generation, high numerical aperture lenses, and near-infrared laser sources \cite{Zumbusch1999}. Over the next several years, CARS microscopy was improved to the point of video-rate \textit{in vivo} imaging \cite{Evans2005a}, but several characteristics limited the utility of the technique: the generation of a nonresonant background (NRB) signal that distorts the resonant, vibrational signal through coherent mixing; the quadratic scaling of the signal with molecular oscillator density, which diminishes or precludes the capacity of CARS to obtain full fingerprint spectra from low-concentration species that are vital in discriminating subtle states of biological systems; and the use of narrowband excitation sources that could only probe narrow regions of the Raman spectrum. The net effect was to limit sensitivity to those species with high oscillator density and strength, and limit specificity to species that possess uniquely isolated Raman peaks. To circumvent the specific challenges posed by the NRB, several methods of background suppression were developed, such as polarization CARS \cite{Cheng2001b} and time-resolved CARS \cite{Volkmer2002}, but with each introducing new technical difficulties and a general reduction of acquisition speed. As an alternative to CARS, stimulated Raman scattering (SRS) was introduced as a CRI technique \cite{Freudiger2008,Ozeki2009} that was NRB-free, capable of high-speed imaging \cite{Saar2010}, and could detect Raman vibrational bands within the fingerprint region \cite{Zhang2012}. Like its CARS counterpart, however, SRS requires multiple acquisitions while varying laser wavelength to address multiple Raman energies \cite{Beier2011,Ozeki2012,Zhang2013}, and rapid wavelength sweeping \cite{Ozeki2012,Zhang2013} is thus far limited to (200 to 300) cm$^{-1}$.

An alternative to swept-source acquisition is the excitation of multiple Raman transitions simultaneously as employed in broadband/multiplex CARS (BCARS/MCARS) \cite{Mueller2002,Cheng2002,Kano2004,Kee2004} and multiplex SRS\cite{Bachler2012,Ploetz2007,Rock2013,Fu2012,Kong2013}. Although a promising new technology, multiplex SRS is currently limited by issues such as laser bandwidth\cite{Fu2012,Rock2013,Bachler2012}, pulse shaping refresh rates \cite{Fu2012,Kong2013}, course spectral resolution\cite{Fu2012}, and interference from competing nonlinear optical phenomena\cite{Ploetz2009}. BCARS also has some drawbacks: the limited spectral energy density of the broadband laser source and slow multielement detection has limited the sensitivity of these system to just the strongest few Raman fingerprint peaks with 10's of millisecond to 100's of millisecond dwell times. On the other hand, BCARS is able to probe the entire spectral range of interest for biological Raman scattering ($>$ 3,000 cm$^{-1}$) with high spectral and spatial resolution. The extended contiguous spectral range of the signal allows for the retrieval of NRB-free Raman spectra through mathematical transforms \cite{Vartiainen1992a,Liu2009,Cicerone2012} obviating the need to experimentally suppress the NRB. Rather than suppress the NRB, we show in this work, that by intentionally generating a strong NRB for use as a local oscillator, we can amplify the weak fingerprint peaks and obtain broadband Raman spectra one to two orders-of-magnitude faster than previously possible and with high spectral clarity. This significant increase in imaging rate and sensitivity to the fingerprint region now positions CRI methods to leverage the many decades of spontaneous Raman work with biological systems, paving a way toward integrating CRI into widespread biological and clinical use.

\section*{Experiments}
\begin{figure}[ht]
\begin{center}
\includegraphics{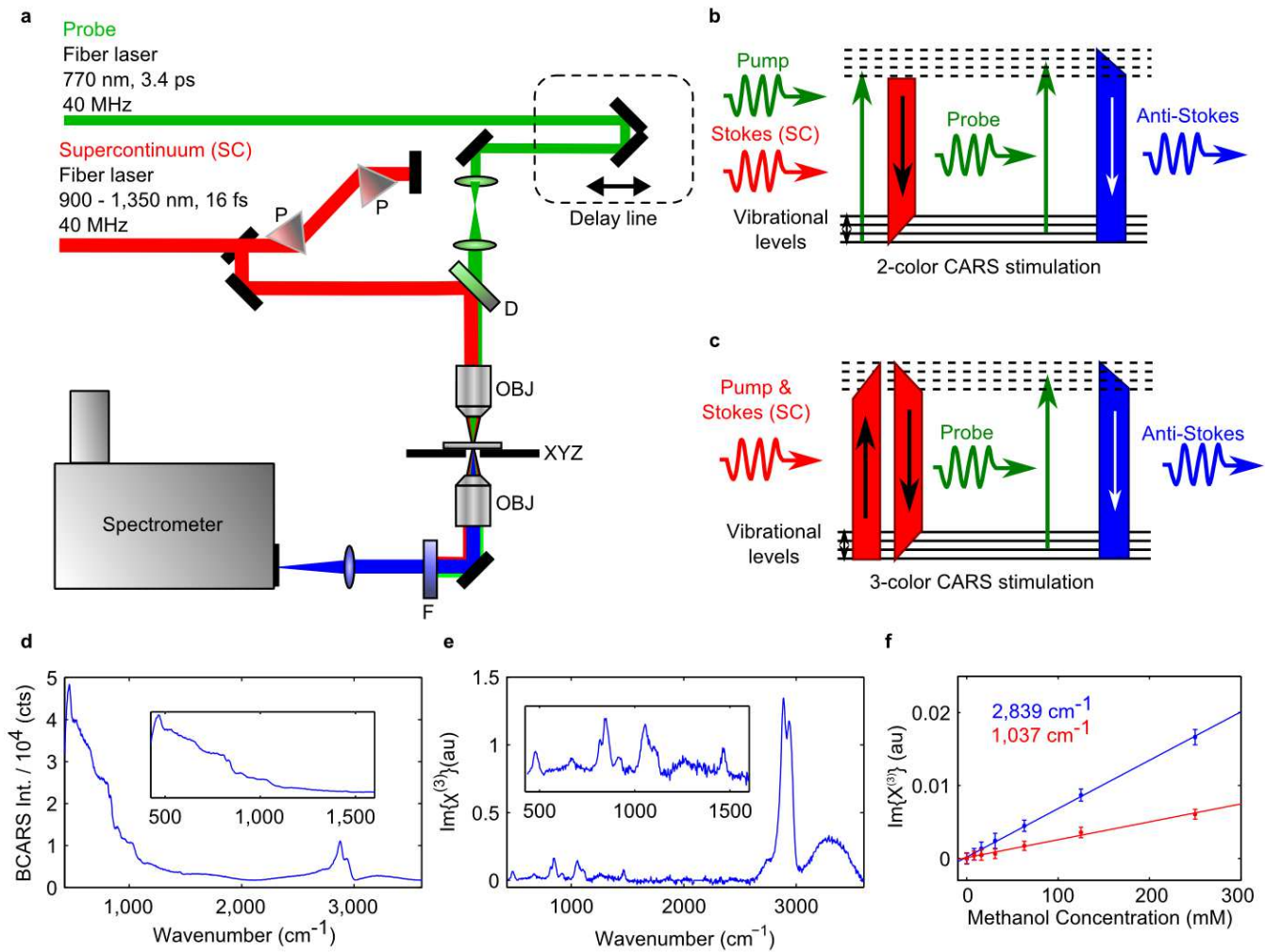}
\caption{\textbf{Coherent Raman Imaging with BCARS microspectroscopy.} \textbf{a,} Schematic of the BCARS CRI system; P, SF10 prism; D, dichroic mirror; OBJ, objective lens; XYZ, piezoelectric stage; F, two short-pass filters. \textbf{b,} Energy diagram with 2-color excitation. \textbf{c,} Energy diagram with 3-color excitation. \textbf{d,} BCARS spectrum of 99 \% glycerol at 3.5 ms exposure. \textbf{e,} Retrieved Raman spectrum of 99 \% glycerol using the Kramers-Kronig transform. \textbf{f,} Linear dependence of the retrieved Raman spectrum on methanol concentration showing a detection limit of $<$ 8 mmol/L using the 2,839 cm$^{-1}$ peak and $<$ 28 mmol/L using the 1,037 cm$^{-1}$ peak. Error bars are $\pm$1 standard deviation.}
\label{Fig1}
\end{center}
\end{figure}

Figure \ref{Fig1}a presents a schematic of the BCARS microspectroscopy system. Laser pulses from two co-seeded fiber lasers \cite{Selm2010} are temporally and spatially overlapped. The probe laser is customized to generate $\approx$ 3.4 ps square-top pulses at 770 nm, providing $<$ 13 cm$^{-1}$ spectral resolution (full width at half maximum). The supercontinuum (SC) laser generates $\approx$ 16 fs pulses spanning ($\approx$ 900 to 1,350) nm with the maximum intensity $\approx$ 1,000 nm. The sources are tailored to excite the strong CH-/OH-stretch region of the Raman spectrum using a ``2-color'' excitation method (see Figure \ref{Fig1}b) \cite{Lee2007}, common to most BCARS/MCARS systems in which the pump and probe sources are degenerate \cite{Mueller2002,Cheng2002,Kano2004,Kee2004}. For detection of the typically weak peaks at low wavenumbers, we excited the fingerprint region with a ``3-color'' excitation profile in which the pump and Stokes transitions are both stimulated by the SC pulse as described in Figure \ref{Fig1}c \cite{Lee2007} (see Supplementary Section `2-Color and 3-Color Excitation Methods'). This method generates the strongest resonant and nonresonant response at the lowest energy levels. And although the NRB limits the vibrational sensitivity and specificity of narrowband CARS techniques \cite{Zumbusch1999}, in spectroscopic CARS techniques, the NRB can be used as a robust local oscillator for heterodyne amplification of the resonant signal. The effect of this approach is to bring the small Raman peaks above the noise floor (see Supplementary Section `Nonresonant Background as Heterodyne Amplifier'). The spectra generated by this combination of 2-color and 3-color excitation are collected with a spectrometer equipped with a thermoelectrically cooled charged-coupled device (CCD) camera that affords acquisition times down to 3.5 ms per spectrum. Additionally, the spectrometer, with particular choice of grating, records a $>$ 250 nm range; thus, we can acquire BCARS spectra and other nonlinear processes, such as second-harmonic generation (SHG) and two-photon excited fluorescence (TPEF), providing an additional layer of information for BCARS spectral interpretation. Figure \ref{Fig1}d shows a raw BCARS spectrum of 99 \% glycerol (acquisition time: 3.5 ms, signal-to-noise ratio [SNR]: 15 dB to 23 dB), which shows the intense 3-color response in the range ($\approx$ 425 to 2,000) cm$^{-1}$ that dwarfs the 2-color response ($\approx$ 2,000 to 3,600) cm$^{-1}$. Although the raw BCARS spectrum is distorted due to coherent mixing between the resonant CARS signal and the NRB \cite{Zumbusch1999}, Figure \ref{Fig1}e demonstrates the use of a time-domain Kramers-Kronig (TDKK) transform to retrieve the imaginary component of the nonlinear susceptibility \cite{Liu2009}, $\Im\{\chi^{(3)}\}$ (convolved with the probe source spectral profile), that is proportional to the [spontaneous] Raman response of the molecule (BCARS peak locations and intensities agree closely with spontaneous Raman experimental results \cite{Mendelovici2000}). We use the TDKK for its speed advantage over competing techniques \cite{Cicerone2012}. To quantify the detection limit of the BCARS microscope and demonstrate molecular response linearity, we recorded spectra from a methanol-water dilution series. As shown in Figure \ref{Fig1}f, the response of the retrieved $\Im\{\chi^{(3)}\}$ is linear with respect to methanol concentration (starting from 1 mol/L; Figure \ref{Fig1}f plotted closer to the dilution limit for clarity), and the detection limit of the system was determined to be $<$ 23 mmol/L using the C-O stretch peak at $\approx$ 1,037 cm$^{-1}$ and $<$ 8 mmol/L using the C-H stretch peak at $\approx$ 2,839 cm$^{-1}$, which matches closely with similar SRS measurements \cite{Freudiger2008}.

\begin{figure}[!ht]
\begin{center}
\includegraphics{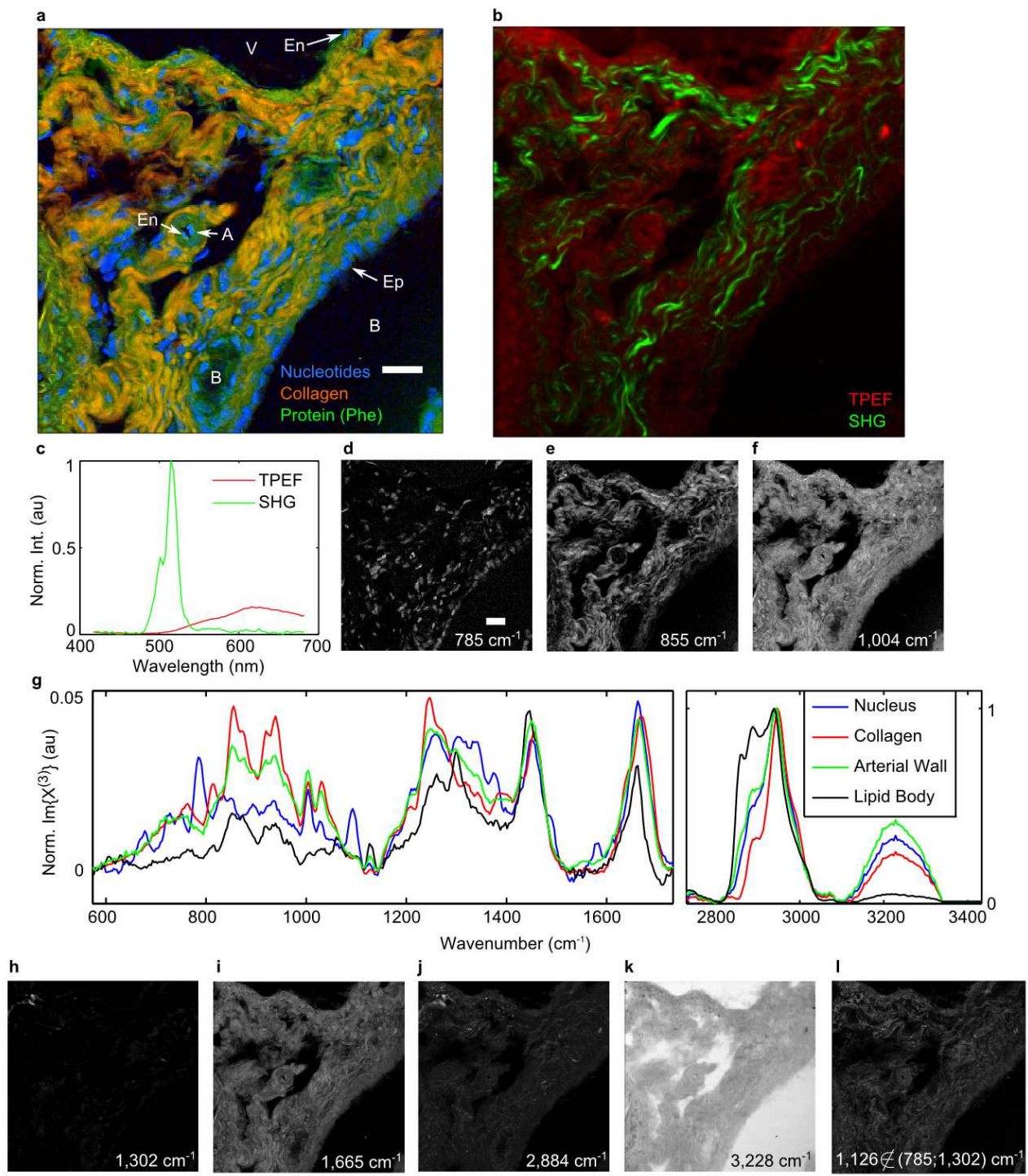}
\caption{\textbf{CRI of murine liver tissue.}  \textbf{a,} Spectral image of portal triad within murine liver tissue with the nuclei in blue, collagen in orange, and protein content in green. A, portal artery; B, bile duct; V, portal vein; Ep, epithelial cell; En, endothelial cell. Scale bar is 20 $\mu$m. \textbf{b,} SHG image highlighting the fibrous collagen network. Scale bar is 20 $\mu$m. \textbf{c,} SHG spectrum for a single pixel. \textbf{d-f,} Spectral images of individual vibrational modes represented by the color channels in a: \textbf{d,} 785 cm$^{-1}$; \textbf{e,} 855 cm$^{-1}$; \textbf{f,} 1,004 cm$^{-1}$. \textbf{g,} Single-pixel spectra from within the nucleus (DNA), collagen fiber, arterial wall, and a lipid droplet. \textbf{h-l,} Additional spectral channels that provide histochemical contrast: \textbf{h,} 1,302 cm$^{-1}$; \textbf{i,} 1,665 cm$^{-1}$; \textbf{j,} 2,884 cm$^{-1}$; \textbf{k,} 3,228 cm$^{-1}$; \textbf{l,} 1,126 cm$^{-1}$ but not containing (785 or 1,302) cm$^{-1}$. Scale is same as in d-f (see scale bar in d).}
\label{Fig2}
\end{center}
\end{figure}

To date, histological analysis of tissues using CRI has relied on limited spectral information primarily in the strong CH-/OH-stretch region of the Raman spectrum ($\approx$ 2,700 cm$^{-1}$ to 3,500 cm$^{-1}$). With these limitations and the complexity of tissue specimens, spectrally identifying and segmenting even such common features as nuclei are nontrivial tasks. This newly developed BCARS platform probes not only the strong CH-/OH-stretch regions of the Raman spectrum, but also the weak and spectrally dense fingerprint region. To demonstrate the sensitivity of the CRI system using molecular fingerprint signatures, we imaged fresh murine liver sectioned to a nominal thickness of 10 $\mu$m. The tissues were mounted in phosphate buffered saline (PBS) between a charged glass slide and a coverslip. Figure \ref{Fig2}a shows a pseudocolor image of liver tissue near a portal triad (hepatic artery, hepatic portal vein, and bile duct) that was collected with 3.5 ms dwell times over a 200 $\mu$m $\times$ 200 $\mu$m area (300 pixels $\times$ 300 pixels). This image contrasts nuclei in blue based on the Raman band $\approx$ 785 cm$^{-1}$, which emanates from DNA/RNA pyrimidine ring breathing and the phosphodiester-stretch \cite{Deng1999}. For further chemical contrast or specificity, one could use other isolated nucleotide peaks at (668, 678, 728, 750, 829, 1,093, 1,488, and 1,580) cm$^{-1}$. Additionally, the peak at 830 cm$^{-1}$, for example, could be used to gauge the amount of DNA in the B-conformation relative to the total genetic content providing information about the functional state of the cells. As a general protein contrast, the ring breathing contribution of phenylalanine at 1,004 cm$^{-1}$ is presented in green. The collagen is highlighted in red using the 855 cm$^{-1}$ C-C-stretch from the pyrrolidine ring of proline (although another C-C stretch peak at 938 cm$^{-1}$ also provides similar contrast) \cite{Frushour1975}.  Previous CRI investigations of tissue incorporated SHG and TPEF imaging to identify collagen and elastin, respectively \cite{Le2007,Meyer2011}, which is shown in Figure \ref{Fig2}b, with demonstrative spectra in Figure \ref{Fig2}c. It should be noted, however, that SHG and TPEF provide uncertain chemical specificity as other biologically-relevant molecular species are known to generate a response as well \cite{Zipfel2003}. Accordingly, we note that areas identifiable as collagen or elastin by SHG or TPEF include species that are clearly not collagen or elastin by their Raman spectra.

With this level of spatial resolution and chemical contrast, several hepatic structures are identifiable by their histology: the hepatic artery with its circular protein-rich, collagen-poor band (likely smooth muscle) surrounding a thin endothelial layer and lumen, the bile ducts lined by tightly packed cuboidal epithelial cells, and the relatively large portal vein with its sparse (due to microtome sample preparation) endothelial layer. Additionally one can see the connective tissue septa (primarily collagen) that enmeshes the portal triad. 

Although the pseudocolor image in Figure \ref{Fig2}a is limited to 3 colors, which are broken down into high-contrast grayscale images in Figures \ref{Fig2}d-f, one can identify significant spectral complexity in the sample as illustrated by the single-pixel spectra in Figure \ref{Fig2}g. Using isolated peaks, one could create dozens of unique images based on vibrational susceptibilities, such as those shown in Figures \ref{Fig2}h-k: 1,302 cm$^{-1}$ (CH$_2$-deformation), 1,665 cm$^{-1}$ (Amide I/C=C-stretch), 2,884 cm$^{-1}$ (CH$_2$-stretch), 3,228 cm$^{-1}$ (O-H-stretch), respectively. Additionally, multivariate analysis of contributions from several peaks, their locations, intensities, and shapes present significant avenues of chemical contrast. For example, Figure \ref{Fig2}l highlights elastin by isolating the chemical species that have vibrations at 1,126 cm$^{-1}$ and lack vibrations at 785 cm$^{-1}$ and 1,302 cm$^{-1}$. The 1,126 cm$^{-1}$ peak is due to the elastin C-N stretch, but also has contributions from lipid C-C stretch, whereas the 1,302 cm$^{-1}$ peak is due primarily to the CH$_2$-deformation of lipids \cite{Frushour1975}. Similarities and differences between the BCARS image and the TPEF image in Figure \ref{Fig2}b, indicate that although elastin is the most abundant fluorescent molecule, multiple contributing chemical species exist.

\begin{figure}[!ht]
\begin{center}
\includegraphics{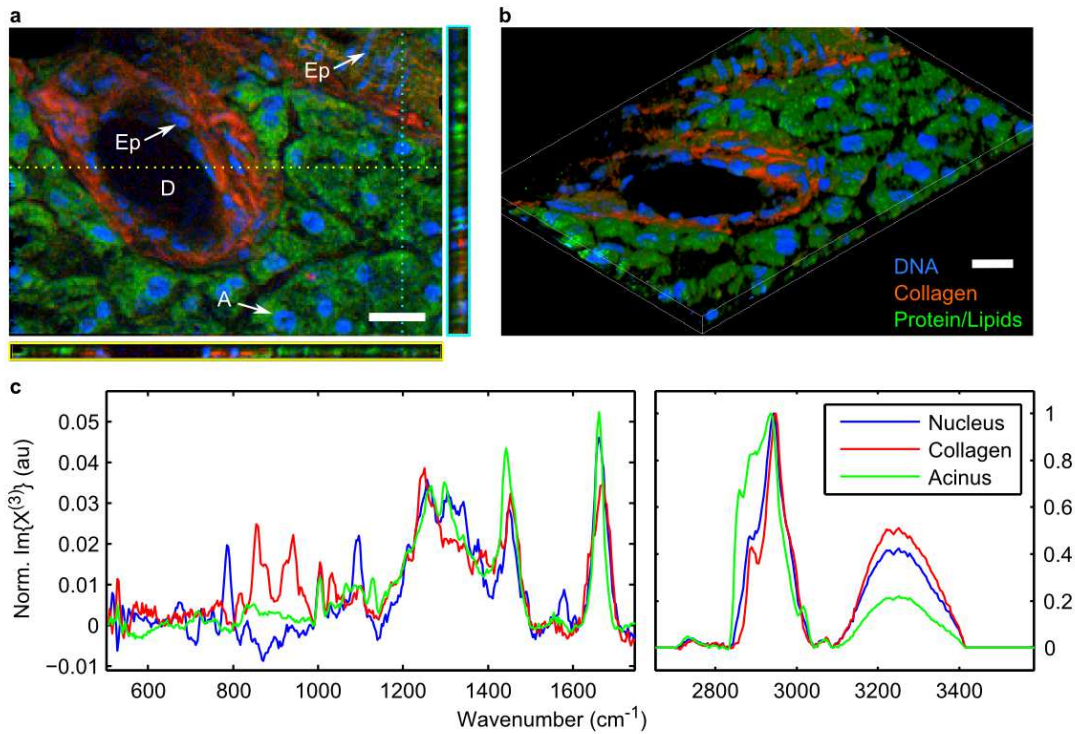}
\caption{\textbf{Three-dimensional CRI of murine pancreatic ducts.}  \textbf{a,} Pseudocolor image taken from a single plane of a Z-stack image collection of exocrine ducts highlighting nuclei (785 cm$^{-1}$) in blue, collagen (855 cm$^{-1}$) in red, and a composite of lipids and proteins in green (1,665 cm$^{-1}$). D, exocrine duct; A, acinar cell; Ep, epithelial cell. Additionally, two axial planes are shown to provide histochemical depth information. Scale bar is 20 $\mu$m. \textbf{b,} Three-dimensional reconstruction of pancreatic ducts from 10 z-stack images. \textbf{c,} Single-pixel spectra taken from within an epithelial cell nucleus, within the fibrous collagen, and from within the cytosol of an acinar cell.}
\label{Fig3}
\end{center}
\end{figure}

Beyond histochemical imaging in two dimensions, the nonlinear excitation of the BCARS system intrinsically limits the axial focal volume; thus, affording the generation of ``z-stack'' images in three dimensions. Narrowband CARS and SRS have demonstrated this capability \cite{Saar2010,Ozeki2012,Ozeki2009,Zumbusch1999}, but three-dimensional microspectroscopy with CRI or spontaneous Raman is uncommon due to long acquisition times. Figure \ref{Fig3}a is a pseudocolor image of murine pancreas highlighting the nuclei in blue (785 cm$^{-1}$), collagen in red (855 cm$^{-1}$), and a general contrast for lipids and protein in green (1,665 cm$^{-1}$: lipids, C=C-stretch; proteins, Amide I). This image shows a single plane from a 10-stack collection with each plane covering 150 $\mu$m $\times$ 100 $\mu$m (0.667 $\mu$m lateral, 1 $\mu$m axial step size; 3.5 ms dwell time; less than 2 minutes per image). Additionally, two reconstructed axial planes are shown. This image shows a small (interlobular) exocrine duct surrounded by epithelial cells, the edge of a larger (interlobular) exocrine duct (as identified by the columnar epithelial cells), a collagen matrix surroundings the ducts, and acinar cells (and the lumen separating the acini). Figure \ref{Fig3}b shows the reconstructed 3D image that more clearly shows the shape, size, and orientation of the individual cells and tissue constituents. Figure \ref{Fig3}c shows single pixel spectra collected from components of the tissue: the nucleus of an epithelial cell, collagen surrounding the small exocrine duct, and from the cytosol of an acinar cell.

\begin{figure}[!ht]
\begin{center}
\includegraphics{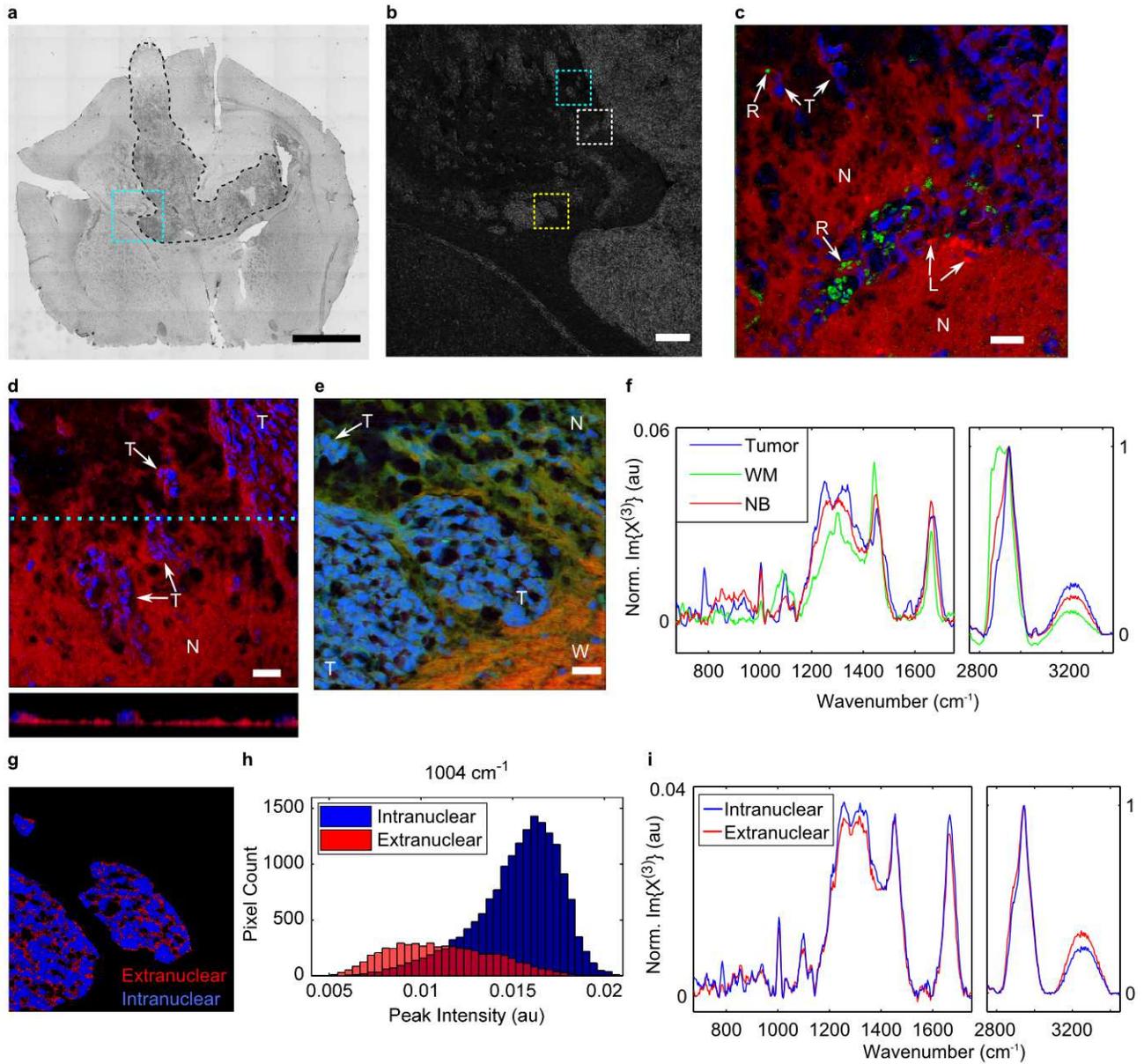}
\caption{\textbf{Histopathology using broadband CRI.}  \textbf{a,} Brightfield image of orthotopic xenograft of human glioblastoma in mouse brain. Hard boundary of tumor is traced in black dashes. Scale bar is 2 mm. \textbf{b,} Phase contrast micrograph of area depicted within cyan dashed-box in subfigure a showing invasive tumor cells. Scale bar is 200 $\mu$m. \textbf{c,} Pseudocolor image of tumor and normal brain tissue highlighting nuclei in blue (730 cm$^{-1}$), lipid content in red (2,850 cm$^{-1}$), and red blood cells in green (1,548 cm$^{-1}$ + 1,566 cm$^{-1}$). This image was collected from the region indicated by the white-dashed box in b. N, normal brain tissue; T, tumor cells; R, red blood cells; L, lipid bodies. Scale bar is 20 $\mu$m. \textbf{d,} Pseudocolor image and axial scan of tumor and normal brain tissue highlighting nuclei in blue (730 cm$^{-1}$) and lipid content in red (2,850 cm$^{-1}$). This image was collected from the region indicated by the cyan-dashed box in b. N, normal brain tissue; T, tumor cells. Scale bar is 20 $\mu$m. \textbf{e,} Pseudocolor image of tumor, white matter, normal brain tissue highlighting nuclei in blue (785 cm$^{-1}$), lipid content in red (2,850 cm$^{-1}$), and CH$_3$-stretch - CH$_2$-stretch (2,944 cm$^{-1}$ - 2,850 cm$^{-1}$) in green.  This image was collected from the region indicated by the yellow-dashed box in b. N, normal brain tissue; T, tumor cells; W, white matter. Scale bar is 20 $\mu$m. \textbf{f,} Single pixel spectra collected from a tumor nuclei, white matter, and normal brain. \textbf{g,} Spectrally segmented pseudocolor image of tumor masses identifying internuclear space (blue) and extranuclear space (red). \textbf{h,} Histogram analysis of phenylalanine content (1,004 cm$^{-1}$) of tumor masses within the intranuclear (blue) and extranuclear (red) regions. \textbf{i,} Mean spectra of intranuclear (blue) and extranuclear (red) space of tumor mass.}
\label{Fig4}
\end{center}
\end{figure}

For histopathological analysis, spontaneous Raman and infrared micro/spectroscopy have experimentally demonstrated adequate chemical specificity and sensitivity to delineate a variety of neoplasms \cite{Huang2003,Haka2005,Gniadecka2004,Koljenovic2002,Nijssen2002,Krafft2010,Kirsch2010,Meyer2011,Krafft2006}, but required long integration times and presented course spatial resolution (typically, 25 $\mu$m to 100 $\mu$m for imaging), which may limit accurate tumor-boundary identification and early-stage tumor detection. Conversely, CRI techniques have demonstrated high-speed, high-spatial resolution imaging of normal and diseased brain tissue but with vibrational contrast limited to single or few energy levels \cite{Meyer2011,Pohling2011,Krafft2010}. We present images of orthotopic xenograft brain tumors within a murine brain with the chemical specificity similar to spontaneous Raman but with speeds afforded by CRI. For imaging, the brain was fixed, frozen, and then sectioned to a nominal thickness of 10 $\mu$m (see Methods for more detail). Figure \ref{Fig4}a shows a brightfield image of a brain slice with an identifiable tumor extending from the cortex to the external boundary. For our experiment, we imaged several areas of the tumor near the center of the brain (see the dashed box in Figure \ref{Fig4}a). Figure \ref{Fig4}b shows a close-up polarization contrast micrograph of the region of interest. Figure \ref{Fig4}c shows a pseudocolor CRI image highlighting the nuclei in blue (730 cm$^{-1}$), lipid-content in red (2,850 cm$^{-1}$), and the red blood cells in green (1,548 cm$^{-1}$ + 1,565 cm$^{-1}$: C-C-stretch from hemoglobin \cite{Wood2002}). This image clearly shows the large tumor mass and a seemingly small projection of neoplastic cells invading healthy brain tissue. Additionally, smaller tumor bodies are identifiable by their high density of distorted nuclei and high nuclear:cytoplasmic ratio that contrasts with the sparse nuclear content of healthy surrounding brain tissue. Furthermore, several lipids bodies are visible at the boundary between the tumor and normal brain tissue. These lipid bodies show an increased response at 2,850 cm$^{-1}$ and a further increase at $\approx$ 701 cm$^{-1}$ and $\approx$ 715 cm$^{-1}$, which may indicate an increase in cholesterol and choline head groups, respectively \cite{Krafft2006,Meyer2011}. Figure \ref{Fig4}d shows a chemical image with the same color scheme as Figure \ref{Fig4}c, showing several small microsatellites or extensions of the main tumor mass invading healthy brain matter. The mesh-like appearance of the healthy tissue is likely an artifact of sectioning and the freeze-thaw cycle due to the difference in tissue density between the tumor and normal brain that becomes apparent after sample thawing (as shown in the axial-scan in Figure \ref{Fig4}d). Figure \ref{Fig4}e shows a chemical image of a boundary between normal brain tissue (likely, gray matter), white matter, and tumor masses with contrast highlighting lipids in red (2,850 cm$^{-1}$, strongest response from the myelinated axons); CH$_3$-stretch - CH$_2$-stretch (2,944 cm$^{-1}$ - 2,850 cm$^{-1}$), a general contrast; and nuclei in blue (785 cm$^{-1}$). The image shows the fibrous-tract appearance of white matter and strands of myelination around clusters of cancer cells. Figure \ref{Fig4}f shows single-pixel spectra from a nucleus within the central tumor mass, the solid white matter region, and the normal brain region. The spectra indicate lipids (as judged from 2,850 cm$^{-1}$) are most concentrated in the white matter and least in the tumor, which agrees with previous chromatographic \cite{Yates1979} and vibrational spectroscopic studies \cite{Krafft2006,Krafft2004}. Additionally, one sees an increase in response from phenylalanine at 1,004 cm$^{-1}$ and an overall reduction in the lipid-protein ratio in tumor cells relative to healthy brain tissues, both of which have been indicated in several Raman studies \cite{Krafft2004,Krafft2010}, but it was not established whether these changes manifested themselves across the tumor or in certain substructures, such as the nuclei. To further analyze the tumor, we spectrally segmented the tumor masses between intracellular regions and extracellular regions as shown in Figure \ref{Fig4}g. Figure \ref{Fig4}h shows a histogram analysis of each pixel within the tumors indicating that the phenylalanine content is more concentrated within the nuclei, which is also indicated in the normalized spectra in Figure \ref{Fig4}i. Additionally the lipid-protein ratio (2,850 cm$^{-1}$ divided by 1,004 cm$^{-1}$) is largest in normal brain matter (14.5), weakest in the intranuclear tumoral space (6.9), and between the two in the extranuclear tumoral space (12.8). 

\section*{Discussion and Conclusion}
In this work, we have presented results from the development of a new CRI platform with an unprecedented combination of speed, sensitivity, spatial resolution, and spectral breadth. This instrument opens up opportunities in histopathology and, we believe, will greatly expand the impact Raman spectroscopy will provide to biomedical researchers and clinicians by providing high spatial and chemical resolution. By way of example, there is a demonstrated need for chemical and spatial resolution in assessing the disease state, grade, and prognosis for glioblastoma multiforme (GBM), an aggressive type of brain cancer. GBM, a grade IV astrocytic tumor has very high lethality and often regrows and spreads after radiotherapy and surgical resection. GBM also presents a histopathological challenge as it is accompanied by a large heterogeneity of morphological and chemical features. Recently, the advent of genomic sequencing and the Cancer Genome Atlas database have revealed several molecular subclassifications of GBM \cite{Dunn2012,Karsy2012} with studies showing a direct correlation between patient response and survival based on treatment tailored to the molecular subtype of GBM. The speed and spectral sensitivity of the BCARS method described here will facilitate  investigation of chemical and morphological heterogeneity between and within tumor grades and better classify GBM molecular subtypes. This new analysis will be based on spatio-chemical information as opposed to current practices of statistical clustering of semi-quantitative polymerase chain reaction (qPCR) or RNA sequencing microarray data.  Previously, Raman spectroscopic differences have been found between glioma tumor grades \cite{Krafft2006}, but discriminating intra-grade molecular subtypes (such as within GBMs) will require spatio-chemical information of higher resolution than has been previously demonstrated. Additionally, this BCARS system offers the resolution and chemical sensitivity to perform time series studies to investigate tumorigenesis, proliferation, and associated pathological trademarks, such as necrosis and angiogenesis.

In addition to the aforementioned biological investigations, there are a number of opportunities for improving the utility of the BCARS system. For example, advanced pulse shaping techniques could be implemented to tailor the NRB generation to particularly enhance the sensitivity over specific spectral ranges, such as the 1,800 cm$^{-1}$ to 2,200 cm$^{-1}$ range, for which our system currently has the least sensitivity. Additional hardware advances such as epidetection of BCARS, as recently demonstrated \cite{Hartshorn2013}, could expand imaging to thick or opaque samples, and in conjunction with endoscope, microprobe, or fiber probe development, could provide an opportunity for \textit{in vivo} imaging as previously demonstrated with Raman microspectroscopy \cite{Kirsch2010} and CRI \cite{Saar2011,Murugkar2010,Balu2010}.

\section*{Methods}
Any mention of commercial products or services is for experimental clarity and does not signify an endorsement or recommendation by the National Institute of Standards and Technology.

\subsection*{\textit{BCARS Microscope}}
The BCARS microspectrometer is constructed from two co-seeded fiber lasers (Toptica, FemtoPro) that provide attosecond-level synchronization with the narrowband probe laser generating $\approx$ 3.4 ps flattop pulses ($\Delta \Omega \,< $ 10$ cm^{-1}$) at 770 nm (40 MHz repetition rate), and the SC source generating $\approx$ 16 fs pulses spanning ($\approx$ 920 nm to 1350 nm) nm (40 MHz repetition rate). The SC beam is directed into an SF10 prism pair pulse compressor to provide (some) chirp control as to maximize the spectral coherence window; thus, maximizing the width of the 3-color CARS response (additional laser tuning and higher-order chirp can move the 2- and 3-color excitation regions to excite, for example, the Raman quiescent region when analyzing deuterated species or cyano groups). The probe beam is directed to a motorized optical delay line to provide temporal control between the two sources. Additionally, the probe beam size is enlarged by a refractive telescope to closely match the beam size to the back aperture of the objective lens. The two beams are combined at a dichroic filter (Omega, 910DCSPXR) and coupled into an inverted microscope (Olympus, IX71). The excitation beams are focused onto the sample using a water-immersion, 60$\times$ (NA = 1.2) objective lens (Olympus, UPlanSApo IR). The sample is mounted on a 3-axis piezo stage (Physik Instrumente, P-545) that provides 200 $\mu$m $\times$ 200 $\mu$m $\times$ 200 $\mu$m movement with sub-micron precision. The excitation light and the generated anti-Stokes photons are collected and collimated with a 60$\times$ objective lens (NA = 0.7) (Olympus, LUCPlanFL N) and passed through two shortpass filters (Semrock, Brightline 770SP; Chroma, HHQ765SP). The remaining anti-Stokes light is focused with an achromatic lens onto the slit of a spectrograph (Acton, SpectroPro 2300i) that is equipped with a CCD camera (Andor, DU970N-FI) for spectral recording. With typical settings, each spectrum is recorded between $\approx$ 470 cm$^{-1}$ to 3,800 cm$^{-1}$ (full spectral range covers a larger region: $\approx$ 268 nm). The camera is directly synchronized with the piezo stage motion controller to allow constant-velocity raster scanning. Each fast-axis line scan is recorded onto the CCD on-board memory and transferred during slow-axis movement. The camera control and acquisition software and the data storage software were developed in-house using Visual C++ and controlled through a custom LabView (National Instruments) interface. The data is processed in MATLAB (Mathworks) through an in-house developed processing suite. Raw spectral data cubes are de-noised using singular value decomposition (SVD) (it should be noted that the average spectrum in Figure \ref{Fig4}i was taken from data that was not de-noised with SVD as averaging effectively reduced the noise level without additional processing), a time-domain Kramers-Kronig transform (TDKK) for spectral phase retrieval\cite{Liu2009}, and baseline detrended. For the TDKK, the estimated NRB signal was collected from either water or glass (slide or coverslip) with the probe delayed to the earliest overlap with the SC, a region in which the NRB dominates the resonant signal; thus, providing a good approximation to the pure NRB. Baseline detrending was performed by manually selecting local minima isolated from Raman peaks \cite{Krafft2006}. In the event that a sample shows regions of mounting media (water or PBS), the fingerprint region below 1,600 cm$^{-1}$ within these areas could be used as a model for the residual background and subtracted. All pseudocolor images, vibrational images, and spectra were generated in MATLAB, and the 3D-reconstructed image in Figure \ref{Fig3}b was generated in ImageJ (NIH). 

\subsection*{\textit{Tissue Sections}}
Fresh murine liver and pancreas tissues were commercially procured (Zyagen). The samples were shipped on dry ice and stored at -80 $^o$C. Prior to imaging, the samples were thawed for 10 minutes, washed twice in PBS to remove debris and residual cutting media. The tissues were kept wet with PBS and a glass coverslip was placed over the sample and sealed with nail polish. 

Glioblastoma cells (GCs) were isolated from primary surgical GBM biopsy specimens in accordance with protocols approved by the Duke University Medical Center or Cleveland Clinic Foundation Institutional Review Boards. In vivo tumor initiation studies were done with BALB/c nu/nu mice under a Cleveland Clinic Foundation Institutional Animal Care and Use Committee-approved protocol. All transplanted mice were maintained for 100 days or until development of neurologic signs, at which point they were euthanized by CO2 asphyxiation. Brains were removed and fixed in 4 \% paraformaldehyde for 24 hours. Following fixation, brains were submerged in 30 \% sucrose as cryoprotectant for an additional 24 hours. Samples were then frozen in optimal cutting temperature compound (OCT) and sectioned on a cryomicrotome to a nominal thickness of 10 $\mu$m. Prior to imaging, samples were thawed, washed with PBS to remove OCT and debris, then covered with a glass coverslip and sealed with nail polish. 


\section*{Acknowledgments}
The authors wish to thank Qiulian Wu, James Hale, and Maksim Sinyuk of Cleveland Clinic for preparing the pathological tissue specimens, and Stephanie Miller of the University of Maryland for preparation of neat chemical specimens. Additionally, C. H. Camp Jr and J. M. Heddleston wish to thank the National Research Council for support through the Research Associate Program (RAP).

\section*{Author contributions}
C.H.C. designed and performed all tissue experiments, analyzed all data, and drafted the manuscript. M.T.C., with assistance from Y.J.L, developed the original concept for the laser design and complimentary 2-color/3-color excitation scheme. C.H.C. constructed the CRI system, modified the laser system, developed the high-speed acquisition software, and developed the processing software. C.H.C., Y.J.L, and C.M.H. developed signal processing methodology and protocols. M.T.C. and Y.J.L. developed the Kramers-Kronig transform used for retrieving the Raman spectral information from the raw BCARS spectra, and C.H.C. developed the parallelized software implementation for high-speed processing. A.R.H.W., J.M.H., J.N.R., and J.D.L. provided materials and/or the tumor sections and provided histopathology insights and direction. J.M.H. provided critical insights into histopathology for neuro-oncology and assisted in performing the tumor section study, as well as contributing to the text of this manuscript. M.T.C. supervised the study.

\section*{Competing financial interests}
The authors declare no competing financial interests.

\end{document}


 
\title{\LARGE{Supplementary Information} \\\vspace{.25in} \Large{High-Speed Coherent Raman Fingerprint Imaging of Biological Tissues}}

\author{Charles H. Camp Jr.}   
\author{Young Jong Lee}
\affiliation{Biosystems and Biomaterials Division, National Institute of Standards and Technology, 100 Bureau Dr, Gaithersburg, MD 20899, USA}
\author{John M. Heddleston}
\affiliation{Semiconductor and Dimensional Metrology Division, National Institute of Standards and Technology, 100 Bureau Dr, Gaithersburg, MD 20899, USA}
\author{Christopher M. Hartshorn}
\affiliation{Biosystems and Biomaterials Division, National Institute of Standards and Technology, 100 Bureau Dr, Gaithersburg, MD 20899, USA}
\author{Angela R. Hight Walker}
\affiliation{Semiconductor and Dimensional Metrology Division, National Institute of Standards and Technology, 100 Bureau Dr, Gaithersburg, MD 20899, USA}
\author{Jeremy N. Rich}  
\affiliation{Department of Stem Cell Biology and Regenerative Medicine, Cleveland Clinic, 9500 Euclid Ave, Cleveland, OH 44195, USA}
\author{Justin D. Lathia}
\affiliation{Department of Cellular and Molecular Medicine, Cleveland Clinic, 9500 Euclid Ave, Cleveland, OH 44195, USA}
\author{Marcus T. Cicerone}
\email{cicerone@nist.gov}
\affiliation{Biosystems and Biomaterials Division, National Institute of Standards and Technology, 100 Bureau Dr, Gaithersburg, MD 20899, USA}

\maketitle

\setcounter{figure}{0}
\makeatletter 
\renewcommand{\thefigure}{S\@arabic\c@figure} 

\section{2-Color and 3-Color Excitation Mechanisms}

The BCARS system uses a hybrid 2-color/3-color approach to excite Raman transitions spanning from the Raman fingerprint region ($<$ 1,800 cm$^{-1}$) to beyond 3,600 cm$^{-1}$. Raman energies above $\approx$ 2,100 cm$^{-1}$ are excited with a 2-color mechanism in which the pump and probe are degenerate-- the excitation mechanism most often used in CARS, MCARS, and BCARS experiments. The fingerprint region, on the other hand, is excited with 3-color excitation. With 3-color excitation, the pump and Stokes sources are degenerate (intrapulse excitation) and the excitation profile is determined by the permutations of available frequencies (energies) of light \cite{Lee2007}. Because the highest number of permutations is for closely spaced frequencies, the 3-color excitation profile increases with decreasing wavenumber. It is for this reason that 3-color excitation is particularly well suited for stimulating Raman transitions within the vibrational fingerprint region. Additionally, the excitation profile of the 3-color mechanism decreases with increasing wavenumber and with practical consideration of the bandwidth afforded by the 16 fs supercontinuum (SC) laser, we designed our system so that as the 3-color excitation profile drops off within the Raman quiescent region ($\approx$ 1,800 cm$^{-1}$ to 2,600 cm$^{-1}$), the 2-color excitation profile emerges, exciting Raman transitions ranging from $\approx$ 2,100 cm$^{-1}$ to 3,600 cm$^{-1}$. With this system utilizing two distinct excitation schemes, it is instructive to analyze the differences between 2-color and 3-color excitation and to understand the effects on the resonant (Raman) and nonresonant signal generation. In this supplemental section, we will highlight 3 key points:
\begin{itemize}
\item With increasing pump and Stokes source bandwidths and fixed average power, the intensity of 3-color excited components rises and 2-color excited components falls
\item The resonant-to-nonresonant signal ratio is primarily tied to probe source excitation parameters, not the pump and Stokes sources, under typical operating conditions
\item The probe source spectral characteristics predominantly determine the spectral resolution of CARS instruments
\end{itemize}
With illumination of these key findings, one can appreciate the power afforded by the presented BCARS system and understand its strength for spectroscopic detection.

The earliest examination of the effects of source bandwidth in CARS microscopy demonstrated a marked increase of NRB generation with increasing source bandwidth but a far more limited increase in the resonant (Raman) component under the assumption of fixed source energies \cite{Cheng2001}. Under this framework, decreasing the source bandwidths stood to improve the chemical sensitivity and specificity for isolated Raman peaks. That work, although presenting a general mathematical model for the CARS generation process, focused on narrowband CARS with single-element detection; 2-color excitation; and pump, Stokes, and probe sources with the same bandwidths. Since the time of this early contribution, other works have presented theoretical and experimental results concerning source characteristics, such as chirp \cite{Knutsen2006a,Andresen2005a,Pestov2008a,Cheng2002}, polarization \cite{Cheng2002,Cheng2001b}, and general pulse shaping paradigms for enhanced excitation efficiency \cite{Oron2003a,Dudovich2002a}. Although the latter two studies implemented a 3-color excitation method (prior to the existence of this naming convention), these works did not address the uniqueness of the excitation mechanism. More recently, a comparison of 2-color and 3-color excitation was presented \cite{Lee2007}, highlighting distinct characteristics of the two (and presenting the naming convention). In this supplementary section, we will expand upon this early comparison in an effort to clarify the mechanistic distinctions. To begin, we will develop a classical model of the CARS mechanism and demonstrate analytical solutions available for certain simplified scenarios, comparing 2-color and 3-color excitation. It should be noted that this classical model does not account for effects as source depletion and saturation of vibrational and electronic states, but under typical operating conditions, it should adequately describe signal generation. Additionally, we will present supporting simulations and experimental results.

For the CARS process, the output intensity, $I_{CARS}(\omega_{as})$, at anti-Stokes frequency $\omega_{as}$ is proportional to the squared-modulus of the 3$^{rd}$-order nonlinear polarization, $P^{(3)}(\omega_{as})$ \cite{Gomez1996}:
\begin{align}
I_{CARS}(\omega_{as}) &\propto |P^{(3)}(\omega_{as})|^2\\
P^{(3)}(\omega_{as})&\propto\iiint\chi^{(3)}(\omega_{as};\omega_p,-\omega_s,\omega_{pr})\times\\\nonumber
&\quad E_p(\omega_p)E^*_s(\omega_s)E_{pr}(\omega_{pr})\delta(\omega_{as}-\omega_p+\omega_s-\omega_{pr})d\omega_{p}d\omega_{s}d\omega_{pr},
\end{align}
where $\chi^{(3)}(\omega_{as};\omega_p,-\omega_s,\omega_{pr})$ is the nonlinear susceptibility; $E_p(\omega_p)$ is the pump electric field; $E_s(\omega_s)$ is the Stokes electric field; $E_{pr}(\omega_{pr})$ is the probe electric field; and $\omega_p$, $\omega_s$, and $\omega_{pr}$ are the pump, Stokes, and probe frequencies, respectively. We will describe the nonlinear susceptibility as an addition of chemically-nonspecific and chemically-specific terms:
\begin{equation}
\chi^{(3)}(\omega_{as};\omega_p,-\omega_s,\omega_{pr}) = \chi_{NR} + \chi_{R}(\omega_{as}) = \chi_{NR} + \sum\limits_{m}\frac{A}{\Omega_m - (\omega_p - \omega_s)-i\Gamma_m},
\end{equation}
where $\chi_R(\omega_{as})$ is the nonlinear susceptibility for the resonant components and  $\chi_{NR}$ is the nonlinear susceptibility for the nonresonant component that generates the nonresonant background (NRB). Within the expansion of $\chi_R(\omega_{as})$, $A_m$, $\Omega_m$, $\Gamma_m$ describe the Lorentzian profile of the m$^{th}$ Raman peak at frequency $\Omega_m$ with half-width $\Gamma_m$. It should be noted that spontaneous Raman spectra are proportional to the imaginary component of the nonlinear susceptibility, $\Im \lbrace \chi^{(3)}\rbrace$ \cite{Hellwarth1979}. For notational clarity, we will abbreviate the form of the nonlinear susceptibility to $\chi^{(3)}(\omega_p - \omega_s)$. Applying this to equation (2) and integrating over the pump frequencies simplifies the description of the nonlinear polarization:
\begin{align}
P^{(3)}(\omega_{as})&\propto \int\chi^{(3)}(\omega_{as}-\omega_{pr})\left [\int E^*_s(\omega_s)E_p(\omega_{as}+\omega_s-\omega_{pr})d\omega_{s}\right ]E_{pr}(\omega_{pr})d\omega_{pr},
\end{align}
which may be written in a more tractable form:
\begin{align}
P^{(3)}(\omega_{as})&\propto \left \lbrace \chi^{(3)}(\omega_{as}) \left [E_s(\omega_{as}) \star E_p(\omega_{as})\right ] \right \rbrace \ast E_{pr}(\omega_{as}),
\end{align}
where `$\star$' and `$\ast$' are the cross-correlation and convolution operations, respectively. From equation (5), in view of the relation in equation (1), we can establish some intuitive insights about the CARS generation process. Firstly, the cross-correlation term describes the energy (frequency) profile available for material excitation. As the cross-correlation term spectrally broadens, so too will the range of Raman transitions that can be stimulated. Secondly, if the cross-correlation term is sufficiently broad (larger than a Raman lineshape), the probe source bandwidth will determine the spectral resolution of the system, i.e., the narrower the probe bandwidth, the narrower the recorded CARS lineshape-- converging to a full-width at half-maximum (FWHM) of $2\Gamma_m$ for the m$^{th}$ peak. Finally, if the cross-correlation term becomes an autocorrelation, as in the case of 3-color excitation, the material excitation profile will be necessarily centered at $\omega_{as} =$ 0 and symmetric about this point; although, actual measurements correspond to $\omega_{as} > $ 0.

If we assume all sources have Gaussian spectral profiles and real envelopes, i.e., are transform limited and temporally centered, we may write the cross-correlation term:
\begin{align}
E_s(\omega)\star E_p(\omega) &= \int E^*_s(\omega')E_p(\omega + \omega')d\omega' \\
&= \frac{E_{s0}E_{p0}\sigma_p\sigma_s\sqrt{2\pi}}{\sqrt{\sigma_p^2+\sigma_s^2}}e^{-{{\frac{(\omega-\omega_{p0}+\omega_{s0})^2}{2(\sigma_p^2 + \sigma_s^2)}}}},
\end{align}
where $E_s(\omega)$ and $E_p(\omega)$ are the Stokes and pump fields with amplitudes $E_{s0}$ and $E_{p0}$, center frequencies $\omega_{s0}$ and $\omega_{p0}$, and full width 1/e-intensities (standard deviations) of $\sigma_s$ and $\sigma_p$, respectively. For the special case of 3-color excitation, in which the pump and Stoke sources are degenerate, the cross-correlation is an autocorrelation:
\begin{align}
E_{p,s}(\omega)\star E_{p,s}(\omega) &= |E_{p0,s0}|^2\sigma_{p,s}\sqrt{\pi}e^{-{{\frac{\omega^2}{4\sigma_{p,s}^2}}}},
\end{align}
where we have noted the degeneracy of the pump and Stokes fields by using $E_{p,s}(\omega)$, with amplitude $E_{s0,p0}$, 1/e-intensity half-width $\sigma_{p,s}$, and frequency offset $\omega_{p0,s0}$ (although all offset-frequency terms cancel out). For a Gaussian field, $A(\omega)$, of the form $A_0 \exp\lbrace-\omega^2 / 2 \sigma^2\rbrace$, the average power, $\mathcal{P}_A$, is proportional to $|A_0|^2 \sigma \sqrt{\pi}$; therefore, from equations (7) and (8):
\begin{align}
E_s(\omega)\star E_p(\omega) &\propto \sqrt{\mathcal{P}_p\mathcal{P}_s} \frac{\sqrt{2\sigma_s\sigma_p}}{\sqrt{ (\sigma_p^2+\sigma_s^2)}}e^{-{{\frac{(\omega-\omega_{p0}+\omega_{s0})^2}{2(\sigma_p^2 + \sigma_s^2)}}}}\\
E_{p,s}(\omega)\star E_{p,s}(\omega) &\propto \mathcal{P}_{p0,s0}e^{-{{\frac{\omega^2}{4\sigma_{p,s}^2}}}}.
\end{align}
For equation (10), the material excitation from 3-color stimulation, the maximum amplitude occurs at $\omega = 0$. Additionally, this maximum amplitude is constant regardless of source bandwidth with fixed average power. At spectral positions shifted from the origin, the material response will increase with source bandwidth. The case for 2-color excitation differs in that the maximum material response will occur at $\omega = \omega_{p0} - \omega_{s0}$, and will decrease with increasing source bandwidth. For the scenario representative of most multispectral CARS experiments, in which the Stoke source is significantly broader than the pump source, $\sigma_s >> \sigma_p$, equation (9) simplifies:
\begin{align}
E_s(\omega)\star E_p(\omega) \rvert_{\sigma_s >> \sigma_p} &\propto \sqrt{\mathcal{P}_p\mathcal{P}_s} \sqrt{\frac{2\sigma_p}{\sigma_s}}e^{-{{\frac{(\omega-\omega_{p0}+\omega_{s0})^2}{2(\sigma_p^2 + \sigma_s^2)}}}}.
\end{align}
Under these conditions, with fixed average power, the 2-color material excitation maximum amplitude will drop $\propto 1/\sqrt{\sigma_s}$.

\begin{figure}[!ht]
\begin{center}
\includegraphics{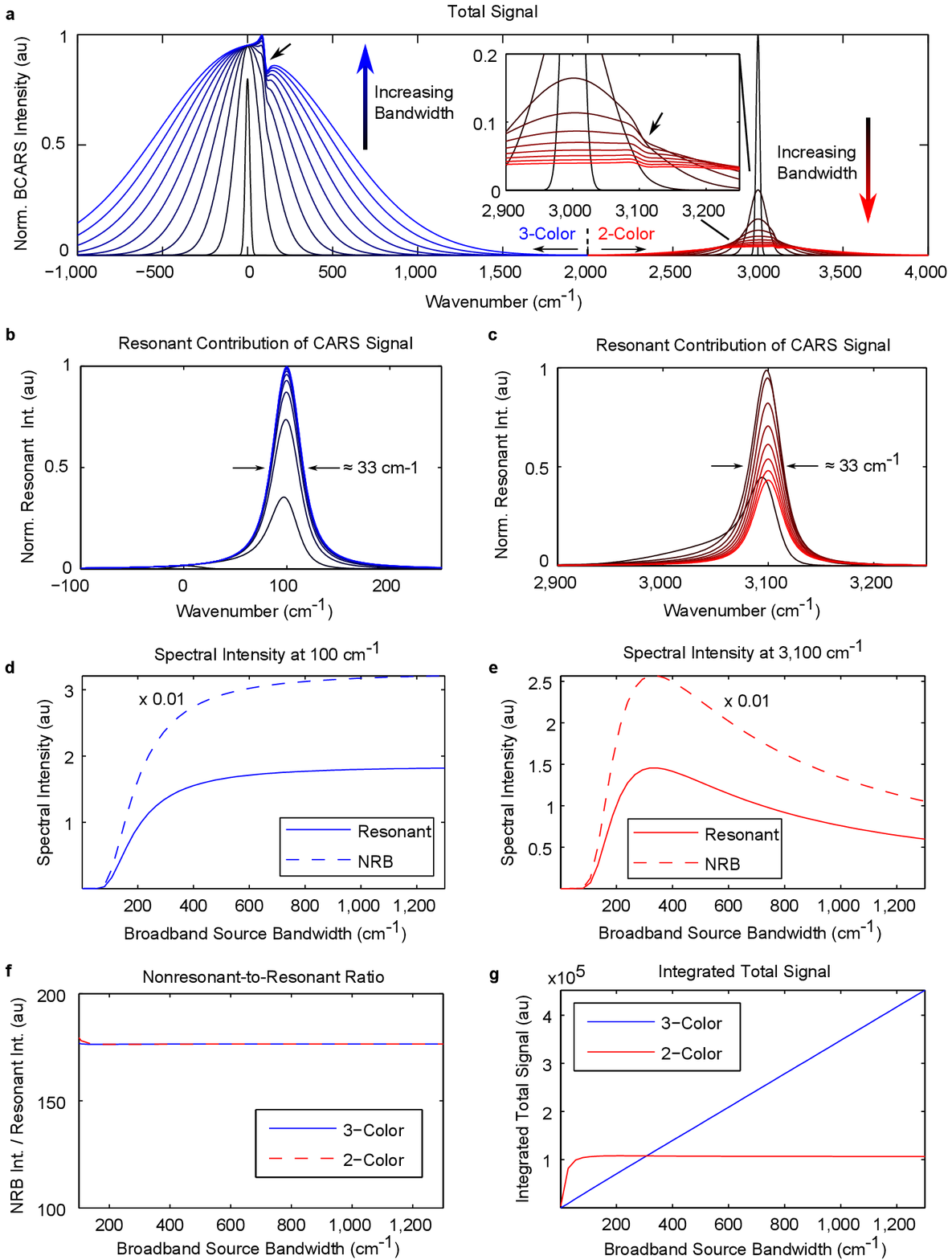}
\caption{\textbf{Simulated effect of SC source bandwidth on NRB generation with a narrowband probe source.} \textbf{a,} Normalized BCARS spectrum from a nonresonant material with 2-color and 3-color excitation mechanisms with varying broadband source bandwidth and fixed average power. \textbf{b,} Resonant component signal from 3-color excitation. \textbf{c,} Resonant component signal from 2-color excitation. \textbf{d,} BCARS signal frequency response for 3-color excitation. \textbf{e,} BCARS signal frequency response for 2-color excitation. \textbf{f,} NRB-to-resonant signal component ratio. \textbf{g,} Integrated total signal.}
\label{FigS1}
\end{center}
\end{figure}

To evaluate these findings and their effects on the total output signal, we will evaluate equation (5) for the case of a nonresonant material:
\begin{align}
I_{CARS}(\omega) &\propto \left | \left \lbrace \chi_{NR} \left [E_s(\omega)\star E_p(\omega) \right ] \right \rbrace \ast E_{pr}(\omega) \right |^2.
\end{align}
Using the cross-correlation and autocorrelation terms in equations (16) and (17) and applying a Gaussian field probe, the 2-color, $I_{2C}(\omega)$, and 3-color, $I_{3C}(\omega)$, CARS signals may be written:
\begin{align}
I_{2C}(\omega) &\propto \left | \frac{2\pi \chi_{NR} E^*_{s0}E_{p0,pr0}^2\sigma_{p,pr}^2\sigma_s}{\sqrt{2\sigma_{p,pr}^2+\sigma_s^2}}e^{-{{\frac{(\omega-2\omega_{p0,pr0}+\omega_{s0})^2}{2(2\sigma_{p,pr}^2 + \sigma_s^2)}}}} \right |^2 \\
&= \frac{4\pi^2 \chi^2_{NR} |E_{s0}|^2 |E_{p0,pr0}|^4 \sigma^4_{p,pr}\sigma^2_s}{2\sigma_{p,pr}^2+\sigma_s^2}e^{-{{\frac{(\omega-2\omega_{p0,pr0}+\omega_{s0})^2}{(2\sigma_{p,pr}^2 + \sigma_s^2)}}}}\\
&\propto \frac{4\sqrt{\pi} \chi^2_{NR} \mathcal{P}_s\mathcal{P}_{p,pr}^2 \sigma^2_{p,pr}\sigma_s}{2\sigma_{p,pr}^2+\sigma_s^2}e^{-{{\frac{(\omega-2\omega_{p0,pr0}+\omega_{s0})^2}{(2\sigma_{p,pr}^2 + \sigma_s^2)}}}}\\
I_{3C}(\omega) &\propto \frac{4\pi^2 \chi^2_{NR} |E_{p0,s0}|^4 |E_{pr0}|^2 \sigma^4_{p,s}\sigma^2_{pr}}{2\sigma_{p,s}^2+\sigma_{pr}^2}e^{-{{\frac{(\omega-\omega_{pr0})^2}{(2\sigma_{p,s}^2 + \sigma_{pr}^2)}}}}\\
&\propto \frac{4\sqrt{\pi} \chi^2_{NR} \mathcal{P}_{p,s}^2 \mathcal{P}_{pr} \sigma^2_{p,s}\sigma_{pr}}{2\sigma_{p,s}^2+\sigma_{pr}^2}e^{-{{\frac{(\omega-\omega_{pr0})^2}{(2\sigma_{p,s}^2 + \sigma_{pr}^2)}}}}.
\end{align}
Comparing these equations, the 2-color maximum signal generation occurs at $\omega = \omega_{p0} + \omega_{pr0} - \omega_{s0}$ and under the condition of a significantly broader Stokes source than pump-probe, the maximum CARS signal will fall $\propto 1/\sigma_s$. For the 3-color excitation case, the maximum signal occurs at frequency $\omega = \omega_{pr}$, and, again, under the case of a relatively narrow probe source and fixed average source powers, the maximum CARS signal will remain constant with increasing pump-Stokes bandwidth. Away from this maximum, with increasing pump-Stokes bandwidth, the CARS signal will rise $\propto \exp\lbrace-(\omega-\omega_{pr0})^2 / (2\sigma_{p,s}^2 + \sigma_{pr}^2) \rbrace$. To demonstrate these findings and to apply equation (5) to the more general case of a material with resonant and nonresonant components, we simulated the CARS signal generation process. In particular, the pump, Stokes, and probe source average powers were held fixed, and the probe intensity FWHM was set to 12 cm$^{-1}$, which is approximately the bandwidth of the probe source in the developed BCARS microscope ($\pm 1$ cm$^{-1}$). For each simulation a single Raman peak (A = 1, 2$\Gamma$ = 20) was simulated +100 cm$^{-1}$ offset from the maximum excitation wavenumber, which for the 3-color case is 0 cm$^{-1}$ and for 2-color stimulation is 3,100 cm$^{-1}$. The nonresonant susceptibility was set to a constant value of 1 ($\max\Im\lbrace\chi^{(3)}\rbrace / \chi_{NR} = 0.1$). Figure \ref{FigS1}a shows the normalized intensity of the CARS spectrum (2-color and 3-color normalized to peak intensity independently) and a variable broadband source. As the theory suggests, with increasing bandwidth, the 3-color regime continues to gain spectral breadth with the response at each spectral component continually rising. Conversely, the 2-color component dramatically decreases in intensity. Figures \ref{FigS1}b and \ref{FigS1}c show the evolution of the resonant component with increasing bandwidth. In the 3-color excitation case, the resonant stimulation rises with increasing bandwidth, as detailed in Figure \ref{FigS1}d. In the 2-color case, the peak intensity at 3,100 cm$^{-1}$ rises as the excitation profile broadens to encompass the entire Raman lineshape. As the excitation profile surpasses this width, the intensity begins to fall, as plotted in Figure \ref{FigS1}e. In both 2-color and 3-color excitation, the spectral resolution, a convolution of the probe source and the Raman nonlinear susceptibility, is the same: $\approx$ 33 cm$^{-1}$. In addition to the resonant signal, Figures \ref{FigS1}d and \ref{FigS1}e describe the strong NRB generation that is $>$ 100$\times$ more intense than the resonant case. In both 2-color and 3-color excitation regimes, the NRB and resonant signal generation follow similar evolution as a function of pump-Stokes source bandwidth. In fact, as shown in Figure \ref{FigS1}f, the nonresonant-to-resonant signal ratios are fixed (standard deviation $\pm$ 0.03 \%) with sufficiently broad excitation sources ($>$ 160 cm$^{-1}$). It should be noted, however, that this ratio can be modulated by changing the probe bandwidth. Finally, if we look at the integrated total signal (resonant and nonresonant contributions), which is proportional to total signal power generated, as shown in Figure \ref{FigS1}g, the 2-color CARS signal is relatively constant with source bandwidths approximately $>$ 160 cm$^{-1}$. 3-color excitation, on the other hand, generates a linearly increasing total signal with increasing source bandwidth. Although it may seem odd that there is an increase in signal with increased bandwidth and that energy is not conserved; in actuality there is just an increase in efficiency. This occurs only in the limit of no excitation source depletion and no vibrational and electronic state saturation.

\begin{figure}[!ht]
\begin{center}
\includegraphics{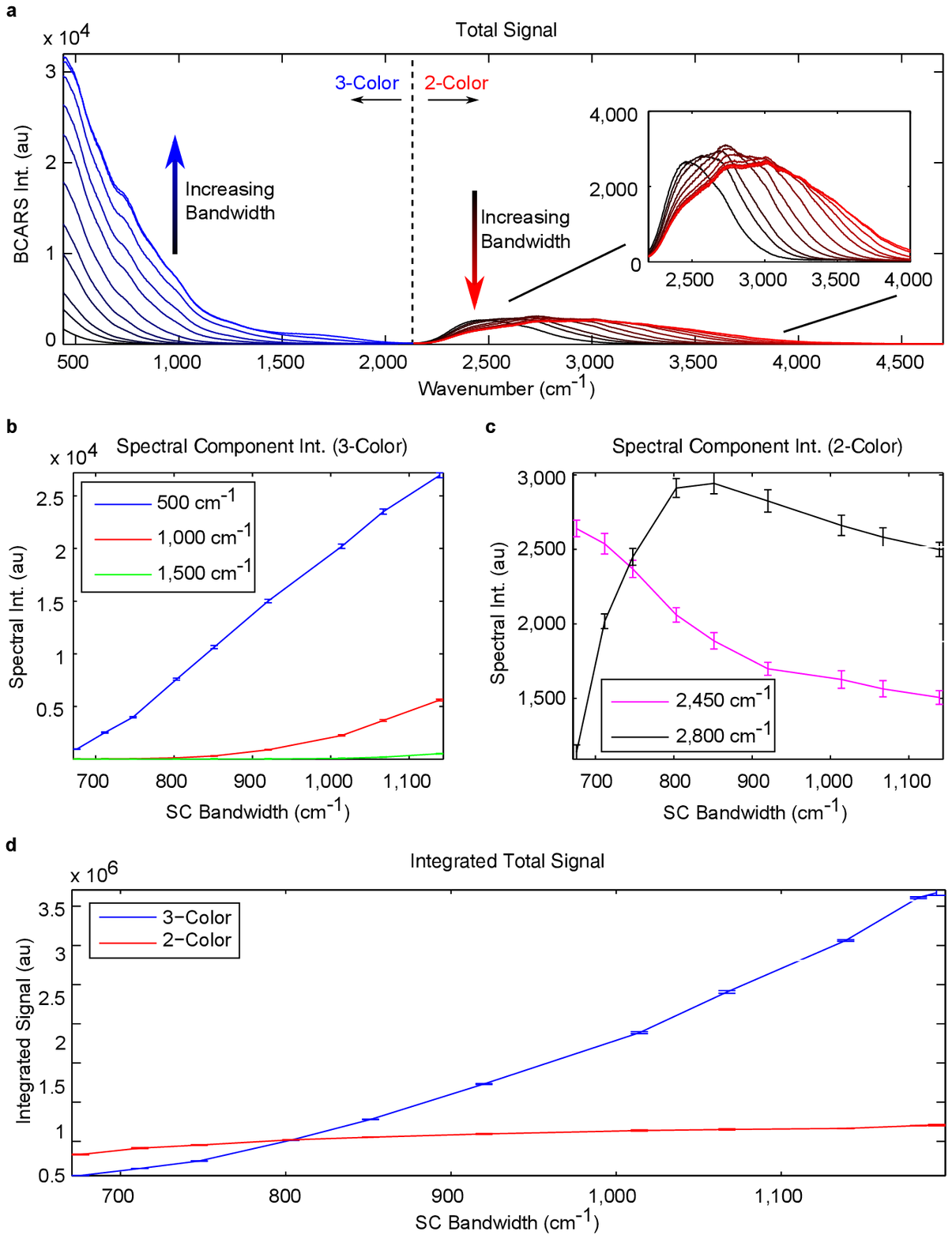}
\caption{\textbf{Experimentally measured effect of SC source bandwidth on NRB generation.} \textbf{a,} BCARS spectrum of glass slide (primarily NRB) with different SC bandwidths. \textbf{b,} BCARS intensity as a function of SC bandwidth for 3-color excitation. \textbf{c,} BCARS intensity as a function of SC bandwidth for 2-color excitation. \textbf{d,} Integrated spectral intensity as a function of SC bandwidth. \textbf{b-d,} Error bars show $\pm$1 standard deviation.}
\label{FigS2}
\end{center}
\end{figure}

To experimentally evaluate some of these simulated and theoretical points, including the counterintuitive behavior of the 3-color signal increase with increased continuum bandwidth, we experimentally measured the BCARS spectrum from a glass microscope slide, which is predominantly NRB. To adjust the bandwidth, a slit was inserted into the prism compressor of the SC source and adjusted to reduce or increase available bandwidth (the slit was inserted on the red-side of the SC spectrum; thus, as bandwidth changes, so too does $\omega_{so}$). A thin-film metal-on-glass attenuator was used to adjust the average power of the SC source to maintain constant average power (6.7 mW), regardless of bandwidth. Figure \ref{FigS2}a shows the recorded BCARS spectra at various SC bandwidths (FWHM). As predicted, with increasing bandwidth, the Raman spectral region excited with 3-color excitation increases universally with increasing source bandwidth. Figure \ref{FigS2}b shows the evolution of signal intensity as a function of source bandwidth at 3 particular wavenumbers. The measured response qualitatively agrees with the simulated results in Figure \ref{FigS1}d. Within the 2-color region, the signal evolves with many similarities to the previously presented simulations. Figure \ref{FigS2}c shows the signal generation at 2 individual frequencies. For 2,450 cm$^{-1}$, which is stimulated under all bandwidth settings, the signal decays with increasing bandwidth. The 2,800 cm$^{-1}$ signal, on the other hand, is not stimulated under narrow source bandwidth conditions; thus, as the slit is opened, the signal rises, plateaus, then decays. Figure \ref{FigS2}d shows the total integrated signal as a function of bandwidth. The 2-color signal is relatively flat and is similar to the findings in Figure \ref{FigS1}g, but does show some increase with increasing bandwidth. Possible explanations are SC source chirp and higher-order dispersion, which were not incorporated into the previous simulations, and source spectral profiles being other than Gaussian. The 3-color signal, conversely, shows a dramatic increase with increasing bandwidth, which also agrees with theoretical and simulated findings. The exact shape of the rise deviates from predictions, and may, too, be partially accounted for by unsimulated SC dispersion and non-Gaussian source spectral envelope shapes. Overall, the experimental results qualitatively agree with the simulations and clearly demonstrate the theoretically predicted trends.

\begin{figure}[!ht]
\begin{center}
\includegraphics{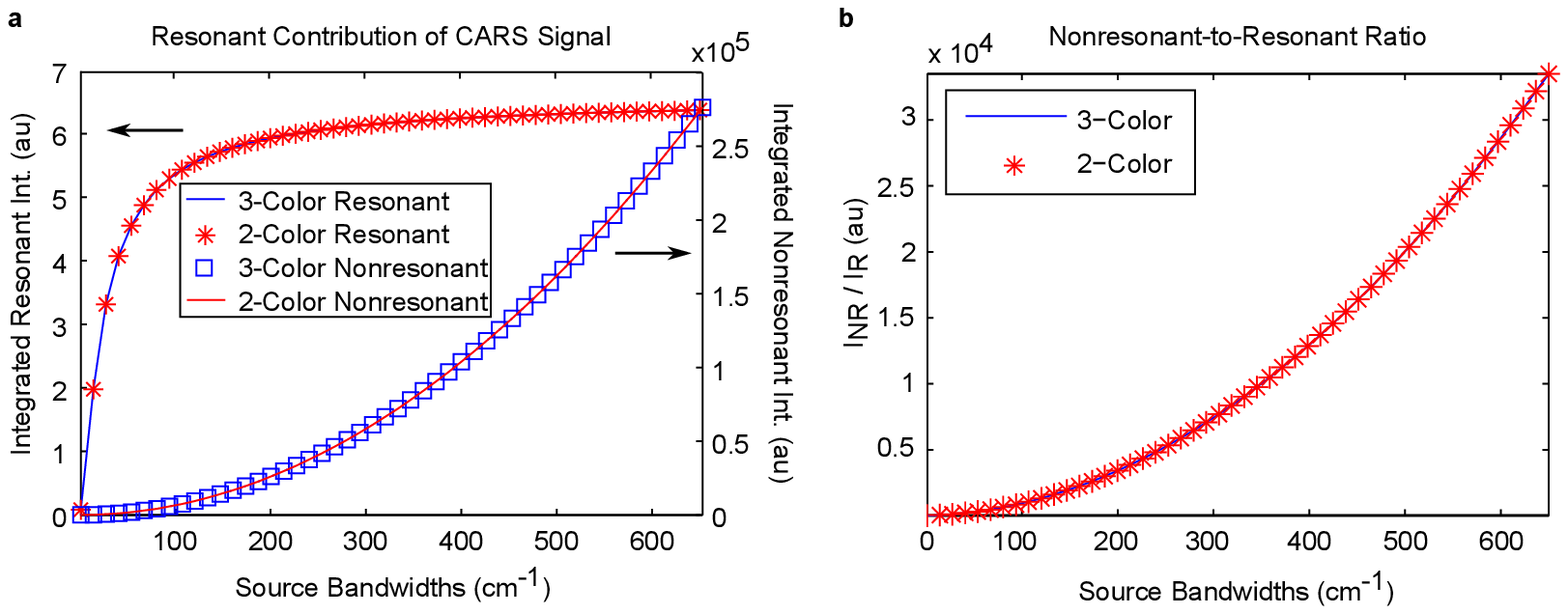}
\caption{\textbf{Simulated CARS signal generation with laser sources of equal bandwidths.} \textbf{a,} Total integrated CARS signal as a function of pump/Stokes/probe source bandwidths. \textbf{b,} Evolution of nonresonant-to-resonant signal component ratio with increasing source bandwidths.}
\label{FigS3}
\end{center}
\end{figure}

In this supplementary section, we described several similarities and differences between 2-color and 3-color excitation. Most importantly we showed that with increasing bandwidth (under the condition of a fixed probe source bandwidth, and fixed average source powers), 3-color excitation response rises, and 2-color excitation signal will eventually decay. Additionally, the theory and simulated results indicated that for excitation sources with stimulation profiles fully encompassing a particular Raman band, the probe source determines the spectral resolution, with the use of an infinitely narrow probe source generating a resonant signal with the same bandwidth as the Raman lineshape. In view of these two characteristics, if the probe source bandwidth, however, were not fixed but varied with the pump and Stokes source bandwidths, as described in early theoretical CARS treatments \cite{Cheng2001}, the resonant and nonresonant signals would evolve differently. Figure \ref{FigS3}a shows the total 2-color and 3-color generated signal as a function of source bandwidths when the pump, Stokes, and probe sources have the same bandwidth under fixed average source powers. Under these conditions with increasing bandwidths, the resonant signal increases but at a decreasing rate, whereas the nonresonant signal increases at an increasing rate. Additionally, if one compares the nonresonant-to-resonant signal ration, as shown in Figure \ref{FigS3}b, the ratio is continually increasing with source bandwidths. Although the generation of NRB is not necessarily an impediment to recovering the resonant signal in spectroscopic CARS techniques, the increasing probe bandwidth will increasingly deteriorate the spectral resolution, smearing out the Raman Lorentzian lineshapes.

\clearpage

\section{The Nonresonant Background (NRB) as Heterodyne Amplifier}

The NRB has often been viewed as an impediment to acquiring resonant CARS signals. In this section, we will theoretically describe and demonstrate with simulations that the NRB can benefit the resonant components of the CARS signal by heterodyne amplification of weak resonant signals above the noise floor of a detector.

For demonstrative purposes, we will consider a hypothetical BCARS system probing a single complex Lorentzian peak:
\begin{align}
I_{CARS}(\omega) &= |\chi_R(\omega)|^2 + |\chi_{NR}(\omega)|^2 + 2\chi_{NR(\omega)}Re\{\chi_R(\omega)\}\\
\chi_R(\omega) &= \frac{A}{\omega - \Omega - i\Gamma}\\
\chi_NR(\omega) &= \chi_{NR}(\omega) \in \Re.
\end{align}
The CARS signal intensity, $I_{CARS}$, is proportional to the squared modulus of the total third-order nonlinear susceptibility convolved with the probe source field, but for simplicity we have assumed a delta function-like probe source and a real, constant value for the cross-correlation of the pump and Stokes sources so that we can encapsulate the pump and Stokes source intensities into the nonlinear susceptibility amplitudes. With these assumptions we define the signal-to-noise ratio (SNR) as:
\begin{align}
SNR(\omega) &= \frac{|\chi_R(\omega)|^2 + 2\chi_{NR}(\omega)Re\{\chi_R(\omega)\}}{\sqrt{|\chi_R(\omega)|^2 + |\chi_{NR}(\omega)|^2+ 2\chi_{NR}(\omega)Re\{\chi_R(\omega)\} + N_R^2}}\\ \nonumber
&= \frac{[A^2+2\chi_{NR}A(\omega - \Omega)]/[(\omega - \Omega)^2 + \Gamma^2]}{\sqrt{[A^2+2\chi_{NR}A(\omega - \Omega)]/[(\omega - \Omega)^2 + \Gamma^2]+\chi_{NR}^2 + N_R^2}},
\end{align}
where $N_R^2$ is the read-out noise of the detector (which includes all non-time dependent noises, such as read noise), and the dark noise is assumed negligible as it is typically orders of magnitude smaller than readout noise and shot noise with modern cooled-CCD cameras over short acquisition times. 

\begin{figure}[htb]
\begin{center}
\includegraphics{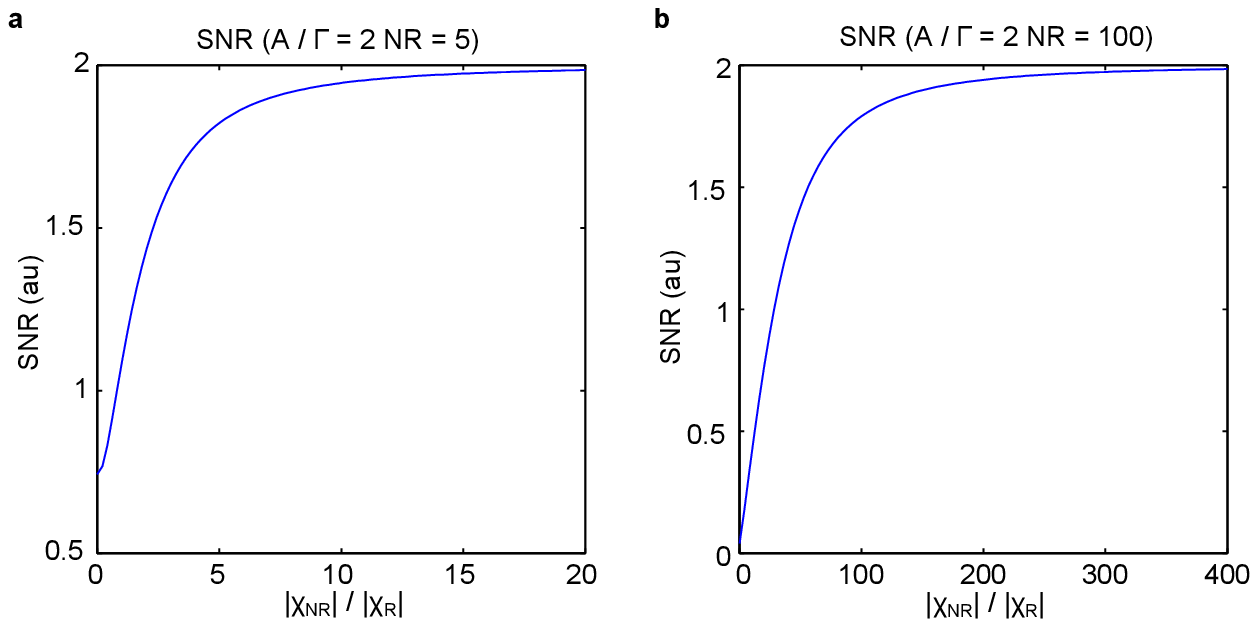}
\caption{\textbf{SNR improvement with increasing NRB.} \textbf{a,} SNR evolution as a function of the nonresonant-to-resonant contribution ratio with a low-noise detector ($N_R=5$) and ,\textbf{b}, a high-noise detector ($N_R = 100$).}
\label{FigS2_1}
\end{center}
\end{figure}

\begin{figure}[htb]
\begin{center}
\includegraphics{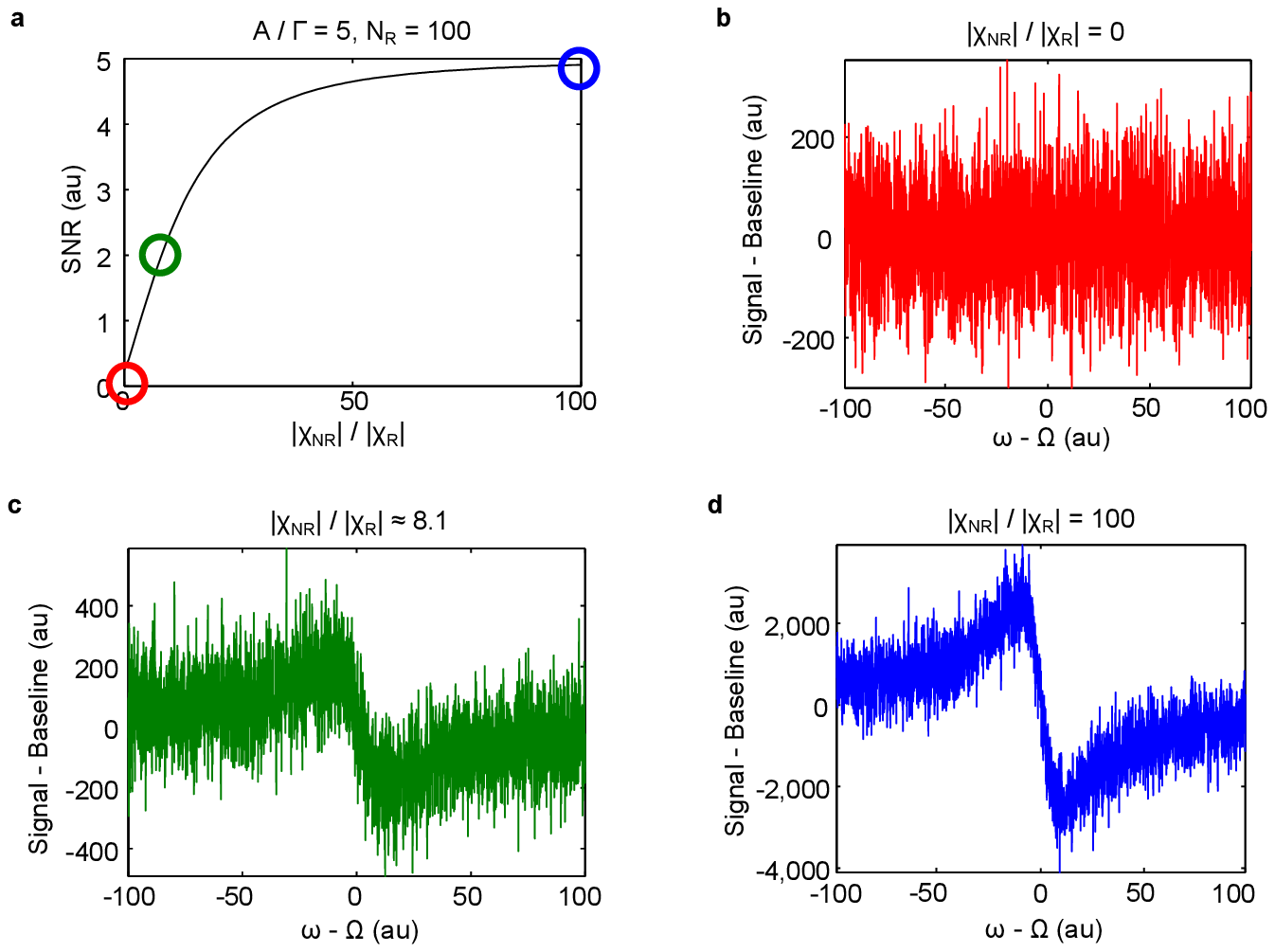}
\caption{\textbf{NRB heterodyne amplification of signal spectrum.} \textbf{a,} SNR evolution as a function of the nonresonant-to-resonant contribution ratio with a high-noise detector ($N_R=100$). The red, green, and blue circles denote the operational conditions for the simulations presented in sub-plots \textbf{b}, \textbf{c}, and \textbf{d}, respectively. \textbf{b,} Simulated spectra with $|\chi_{NR}|/|\chi_{R}|= 0$, $SNR \approx 0.25$. \textbf{c,} Simulated spectra with $|\chi_{NR}|/|\chi_{R}| \approx 8.1$, $SNR \approx 2$. \textbf{d,} Simulated spectra with $|\chi_{NR}|/|\chi_{R}| = 100$, $SNR \approx 4.9$.}
\label{FigS2_2}
\end{center}
\end{figure}

The relative amplitude of the resonant and nonresonant components strongly influences the role that the NRB has on the SNR. Under the condition that the resonant component of the signal is significantly larger than the nonresonant component, the maximum SNR is:
\begin{align}
&SNR(\omega=\Omega) = \frac{A^2/\Gamma^2}{\sqrt{[A^2/\Gamma^2+\chi_{NR}^2 + N_R^2}}\\
&\lim\limits_{A^2/\Gamma^2 \rightarrow \infty} SNR(\omega=\Gamma) = \frac{A}{\Gamma};
\end{align}
thus, the nonresonant component only contributes noise to the signal (i.e., the SNR falls with increasing nonresonant contribution). Practically, this condition may be met in spectral regions of high oscillator density such as within the CH-stretch region of the Raman spectrum, where the resonant component is extremely strong. In other spectral regions, the nonresonant component has a different effect on SNR. For $\chi_R^2 << \chi_{NR}^2$ and at the maximum of the dispersed Raman lineshape ( $\omega=\Gamma+\Omega$), the SNR is given as: 
\begin{align}
&SNR(\omega=\Gamma+\Omega) = \frac{2\chi_{NR}A\Gamma/2\Gamma^2}{\sqrt{2\chi_{NR}A\Gamma/2\Gamma^2+\chi_{NR}^2 + N_R^2}} = \frac{\chi_{NR}A/\Gamma}{\sqrt{\chi_{NR}A/\Gamma+\chi_{NR}^2 + N_R^2}}\\
&\lim\limits_{\chi_{NR}^2 \rightarrow \infty} SNR(\omega=\Gamma+\Omega) = \frac{A}{\Gamma}.
\end{align}
This relationship demonstrates that a large NRB may amplify the signal above the readout noise level when necessary. Additionally, in both the large resonant signal limit and the small resonant signal limit, the SNR asymptotically approaches $A\: / \:\Gamma$. Figure \ref{FigS2_1} shows the calculated SNR for a system with $A\:/\:\Gamma = 2$ for (a) a low-noise detector (similar to many cooled scientific CCD cameras) and (b) a higher-noise detector. In both cases, the SNR with no nonresonant contribution is below one, but with an increasing nonresonant-to-resonant ratio, the maximum SNR approaches $A\:/\:\Gamma$; ergo, the NRB does not affect signal quality. Under these conditions, capturing the CARS signal requires a relatively large NRB. This is further demonstrated in Figure \ref{FigS2_2} for a system with $A\:/\:\Gamma = 5$ with a noisy detector. Figure \ref{FigS2_2}a shows the calculated SNR plot beginning with an SNR below 1 but increasing towards 5. Figures \ref{FigS2_2}b-d show the CARS spectrum for a single simulated peak with noise contributions from the shot noise (both resonant and nonresonant contributions) and readout noise. In \ref{FigS2_2}b, the case of no NRB, the SNR is $\approx 0.25$; thus, the signal is completely masked. In the other extreme with the NRB being 100 times larger than the maximum resonant contribution, the SNR approaches 5 and the signal is easily discernible.
\clearpage